\documentclass{article}

\begin{document}

\centerline{\large\bf The Simple Criteria of SLOCC Equivalence Classes
\footnote{The paper was supported by NSFC(Grant No. 60433050), the basic research fund
of Tsinghua university No: JC2003043 and partially by the state key lab. of
intelligence technology and system}} 
\centerline{Dafa Li$^{a}$\footnote{email address:dli@math.tsinghua.edu.cn},
Xiangrong Li$^{b}$, Hongtao Huang$^{c}$, Xinxin Li$^{d}$ }

\centerline{$^a$ Dept of mathematical sciences, Tsinghua University, Beijing
100084 CHINA}

\centerline{$^b$ Department of Mathematics, University of California, Irvine, CA
92697-3875, USA}

\centerline{$^c$ Electrical Engineering and Computer Science Department} %
\centerline{ University of Michigan, Ann Arbor, MI 48109, USA}

\centerline{$^d$ Dept. of computer science, Wayne State University, Detroit, MI 48202,
USA}

\textbf{Abstract}

We put forward an alternative approach to the SLOCC classification of
entanglement states of three-qubit and four-qubit systems. By directly
solving matrix equations, we obtain the relations satisfied by the
amplitudes of states. The relations are readily tested since in them only
addition, subtraction and multiplication occur.



\eject

\section{Introduction}

Entanglement plays a key role in quantum computing and quantum information
theory. One of the interesting issues on entanglement is how to define the
equivalence of two entangled states. If two states can be obtained from each
other by means of local operations and classical communication (LOCC) with
nonzero probability, we say that two states have the same kind of
entanglement\cite{Bennett}. It is well known that a pure entangled state of
two-qubits can be locally transformed into a EPR state. For multipartite
systems, there are several inequivalent forms of entanglement under
asymptotic LOCC\cite{Bennett2}. Recently, many authors investigated the
equivalence classes of three-qubit states specified by stochastic local
operations and classical communication (SLOCC)\cite{Acin}$-$\cite{Pan}. D%
\"{u}r et al showed that for pure states of three-qubits there are total six
different classes of the entanglement and out of the six classes, there are
two inequivalent types of genuine tripartite entanglement\cite{Dur}. They
put forward a principled method to distinguish the six classes from each
other by calculating the ranks of the reduced density matrices and the
minimal product decomposition\cite{Dur}. For example, they pointed out that
if a state of a three-qubit system with $r(\rho _{A})=r(\rho _{B})=r(\rho
_{C})=2$ has 2 (resp. 3) product terms in its minimal product decomposition
under SLOCC, then the state is equivalent to $|GHZ\rangle $ (resp. $%
|W\rangle $). However, so far no criterion is proposed for the minimal
number of product decomposition terms under SLOCC\cite{Acin}\cite{Dur}\cite%
{Sampera}. In a more recent paper, Verstraete et al\cite{Moor2} considered
the entanglement class of four-qubits under SLOCC and concluded that there
exist nine families of states corresponding to nine different ways of
entanglement. In these previous papers, the authors just put forward some
principled rules of classifying the entangled states. It needs complicated
calculations when these principled rules is used to real states. It will be
quite useful if a feasible approach can be found. Here, we present an
alternative approach to classify the entanglement of three-qubits, and then
generalize the case of four-qubits. We will give simple criteria of
distinguishing the entanglement classes from each other simply by checking
the relations satisfied by the amplitudes of the states.

\section{Classification of entanglement for a three-qubit system}

We first discuss the system comprising three qubits A, B, C. The states of a
three-qubit system can be generally expressed as 
\[
|\psi \rangle =a_{0}|000\rangle +a_{1}|001\rangle +a_{2}|010\rangle
+a_{3}|011\rangle +a_{4}|100\rangle +a_{5}|101\rangle +a_{6}|110\rangle
+a_{7}|111\rangle . 
\]%
Two states $|\psi \rangle $ and $|\psi ^{\prime }\rangle $, are equivalent
under SLOCC if and only if there exist invertible local operators $\alpha
,\beta $ and $\gamma $ such that

\begin{eqnarray}
|\psi \rangle =\alpha \otimes \beta \otimes \gamma |\psi ^{\prime }\rangle ,
\label{psi'}
\end{eqnarray}

where the local operators $\alpha ,\beta $ and $\gamma $ can be expressed as 
$2\times 2$ invertible matrices 
\[
\alpha =\left( 
\begin{array}{cc}
\alpha _{1} & \alpha _{2} \\ 
\alpha _{3} & \alpha _{4}%
\end{array}%
\right) ,~~~~~\beta =\left( 
\begin{array}{cc}
\beta _{1} & \beta _{2} \\ 
\beta _{3} & \beta _{4}%
\end{array}%
\right) ,~\ \ ~~\gamma =\left( 
\begin{tabular}{cc}
$\ \gamma _{1}$ & $\ \gamma _{2}$ \\ 
$\ \gamma _{3}$ & $\ \gamma _{4}$%
\end{tabular}%
\right) . 
\]
We consider the following six classes, respectively.

\subsection{The class equivalent to the state $|GHZ\rangle$}

Let $|\psi ^{\prime }\rangle \equiv |GHZ\rangle $, \textsl{i.e.} 
\begin{eqnarray}
|\psi ^{\prime }\rangle =\frac{1}{\sqrt{2}}(|000\rangle +|111\rangle ).
\label{ghz}
\end{eqnarray}
Substituting Eq. (\ref{ghz}) into Eq. (\ref{psi'}), we get 
\begin{eqnarray}
&&a_{0}=(\alpha _{1}\beta _{1}\gamma _{1}+\alpha _{2}\beta _{2}\gamma _{2})/%
\sqrt{2},~~~~a_{1}=(\alpha _{1}\beta _{1}\gamma _{3}+\alpha _{2}\beta
_{2}\gamma _{4})/\sqrt{2},  \nonumber \\
&&a_{2}=(\alpha _{1}\beta _{3}\gamma _{1}+\alpha _{2}\beta _{4}\gamma _{2})/%
\sqrt{2},~~~~a_{3}=(\alpha _{1}\beta _{3}\gamma _{3}+\alpha _{2}\beta
_{4}\gamma _{4})/\sqrt{2},  \nonumber \\
&&a_{4}=(\alpha _{3}\beta _{1}\gamma _{1}+\alpha _{4}\beta _{2}\gamma _{2})/%
\sqrt{2},~~~~a_{5}=(\alpha _{3}\beta _{1}\gamma _{3}+\alpha _{4}\beta
_{2}\gamma _{4})/\sqrt{2},  \nonumber \\
&&a_{6}=(\alpha _{3}\beta _{3}\gamma _{1}+\alpha _{4}\beta _{4}\gamma _{2})/%
\sqrt{2},~~~~a_{7}=(\alpha _{3}\beta _{3}\gamma _{3}+\alpha _{4}\beta
_{4}\gamma _{4})/\sqrt{2}.  \nonumber
\end{eqnarray}%
By calculating $a_{i}a_{j}-a_{k}a_{l}$, where $i+j=k+l$ and $i\oplus
j=k\oplus l$, where $\oplus $ is addition modulo two,\ we obtain the
following equations: 
\begin{eqnarray}
&&a_{2}a_{4}-a_{0}a_{6}=\allowbreak \gamma _{2}\gamma _{1}\left( \alpha
_{1}\alpha _{4}-\alpha _{3}\alpha _{2}\right) \left( \beta _{2}\beta
_{3}-\beta _{4}\beta _{1}\right) /2,  \nonumber \\
&&a_{3}a_{5}-a_{1}a_{7}=\allowbreak \gamma _{4}\gamma _{3}\left( \alpha
_{1}\alpha _{4}-\alpha _{3}\alpha _{2}\right) \left( \beta _{2}\beta
_{3}-\beta _{4}\beta _{1}\right) /2,  \nonumber \\
&&a_{0}a_{7}-a_{3}a_{4}=\allowbreak -\left( \alpha _{1}\alpha _{4}-\alpha
_{2}\alpha _{3}\right) \left( -\gamma _{4}\beta _{4}\gamma _{1}\beta
_{1}+\beta _{3}\gamma _{3}\beta _{2}\gamma _{2}\right) \allowbreak /2, 
\nonumber \\
&&a_{1}a_{6}-a_{2}a_{5}=\allowbreak -\left( -\alpha _{3}\alpha _{2}+\alpha
_{1}\alpha _{4}\right) \left( -\gamma _{2}\beta _{4}\gamma _{3}\beta
_{1}+\beta _{3}\gamma _{1}\beta _{2}\gamma _{4}\right) /2\allowbreak . 
\nonumber
\end{eqnarray}%
By using the above equations, we further obtain 
\begin{eqnarray}
&&(a_{0}a_{7}-a_{2}a_{5}+a_{1}a_{6}-a_{3}a_{4})^{2}-4(a_{2}a_{4}-a_{0}a_{6})(a_{3}a_{5}-a_{1}a_{7})
\nonumber \\
&=&\frac{1}{4}\left( \alpha _{1}\alpha _{4}-\alpha _{2}\alpha _{3}\right)
^{2}\left( -\gamma _{4}\gamma _{1}+\gamma _{3}\gamma _{2}\right) ^{2}\left(
-\beta _{4}\beta _{1}+\beta _{2}\beta _{3}\right) ^{2}.  \label{ghz2}
\end{eqnarray}%
$|\psi \rangle $ is equivalent to $|GHZ\rangle $ under SLOCC, if and only if
the invertible operators $\alpha ,\beta $ and $\gamma $ exist. From Eq. (\ref%
{ghz2}), we may immediately conclude that the necessary and sufficient
condition of $|\psi \rangle $ being equivalent to $|GHZ\rangle $ is 
\begin{eqnarray}
(a_{0}a_{7}-a_{2}a_{5}+a_{1}a_{6}-a_{3}a_{4})^{2}-4(a_{2}a_{4}-a_{0}a_{6})(a_{3}a_{5}-a_{1}a_{7})\neq 0.
\label{gz3}
\end{eqnarray}

It is not hard to verify that 
\begin{eqnarray}
&&(a_{0}a_{7}-a_{2}a_{5}+(a_{1}a_{6}-a_{3}a_{4}))^{2}-4(a_{2}a_{4}-a_{0}a_{6})(a_{3}a_{5}-a_{1}a_{7})=
\nonumber \\
&&(a_{0}a_{7}-a_{3}a_{4}-(a_{1}a_{6}-a_{2}a_{5}))^{2}-4(a_{1}a_{4}-a_{0}a_{5})(a_{3}a_{6}-a_{2}a_{7})=
\nonumber \\
&&(a_{0}a_{7}-a_{2}a_{5}-(a_{1}a_{6}-a_{3}a_{4}))^{2}-4(a_{0}a_{3}-a_{1}a_{2})(a_{4}a_{7}-a_{5}a_{6}).
\label{ghz4}
\end{eqnarray}
\noindent Therefore the above condition (\ref{ghz3}) can be replaced by the
following any one of the following conditions.

\begin{eqnarray}
(a_{0}a_{7}-a_{3}a_{4}-(a_{1}a_{6}-a_{2}a_{5}))^{2}-4(a_{1}a_{4}-a_{0}a_{5})(a_{3}a_{6}-a_{2}a_{7}) &\neq &0,
\nonumber \\
(a_{0}a_{7}-a_{2}a_{5}-(a_{1}a_{6}-a_{3}a_{4}))^{2}-4(a_{0}a_{3}-a_{1}a_{2})(a_{4}a_{7}-a_{5}a_{6}) &\neq &0.
\end{eqnarray}

\subsection{The class equivalent to the state $|W\rangle $}

Let $|\psi ^{\prime }\rangle \equiv |W\rangle $, \textsl{i.e.} 
\begin{eqnarray}
|\psi ^{\prime }\rangle=\frac{1}{\sqrt{3}}(|001\rangle +|010\rangle
+|100\rangle).  \label{w1}
\end{eqnarray}

Substituting Eq. (\ref{w1}) into Eq. (\ref{psi'}), we get 
\begin{eqnarray}
&&a_{0}=(\alpha _{1}\beta _{1}\gamma _{2}+\alpha _{1}\beta _{2}\gamma
_{1}+\alpha _{2}\beta _{1}\gamma _{1})/\sqrt{3},~~~~a_{1}=(\alpha _{1}\beta
_{1}\gamma _{4}+\alpha _{1}\beta _{2}\gamma _{3}+\alpha _{2}\beta _{1}\gamma
_{3})/\sqrt{3},  \nonumber \\
&&a_{2}=(\alpha _{1}\beta _{3}\gamma _{2}+\alpha _{1}\beta _{4}\gamma
_{1}+\alpha _{2}\beta _{3}\gamma _{1})/\sqrt{3},~~~~a_{3}=(\alpha _{1}\beta
_{3}\gamma _{4}+\alpha _{1}\beta _{4}\gamma _{3}+\alpha _{2}\beta _{3}\gamma
_{3})/\sqrt{3},  \nonumber \\
&&a_{4}=(\alpha _{3}\beta _{1}\gamma _{2}+\alpha _{3}\beta _{2}\gamma
_{1}+\alpha _{4}\beta _{1}\gamma _{1})/\sqrt{3},~~~~a_{5}=(\alpha _{3}\beta
_{1}\gamma _{4}+\alpha _{3}\beta _{2}\gamma _{3}+\alpha _{4}\beta _{1}\gamma
_{3})/\sqrt{3},  \nonumber \\
&&a_{6}=(\alpha _{3}\beta _{3}\gamma _{2}+\alpha _{3}\beta _{4}\gamma
_{1}+\alpha _{4}\beta _{3}\gamma _{1})/\sqrt{3},a_{7}=(\alpha _{3}\beta
_{3}\gamma _{4}+\alpha _{3}\beta _{4}\gamma _{3}+\alpha _{4}\beta _{3}\gamma
_{3})/\sqrt{3}.  \nonumber \\
&&
\end{eqnarray}%
By calculating $a_{i}a_{j}-a_{k}a_{l}$, where $i+j=k+l$ and $i\oplus
j=k\oplus l$, we have 
\begin{eqnarray}
&&a_{0}a_{3}-a_{1}a_{2}=\allowbreak \alpha _{1}^{2}\left( -\beta _{2}\beta
_{3}+\beta _{1}\beta _{4}\right) \left( \gamma _{2}\gamma _{3}-\gamma
_{4}\gamma _{1}\right) /3,  \nonumber \\
&&a_{5}a_{6}-a_{4}a_{7}=\allowbreak -\alpha _{3}^{2}\left( -\beta _{2}\beta
_{3}+\beta _{1}\beta _{4}\right) \left( \gamma _{2}\gamma _{3}-\gamma
_{4}\gamma _{1}\right) /3,  \nonumber \\
&&a_{1}a_{4}-a_{0}a_{5}=\allowbreak -\beta _{1}^{2}\left( \gamma _{2}\gamma
_{3}-\gamma _{4}\gamma _{1}\right) \left( \alpha _{1}\alpha _{4}-\alpha
_{3}\alpha _{2}\right) /3,  \nonumber \\
&&a_{3}a_{6}-a_{2}a_{7}=\allowbreak -\beta _{3}^{2}\left( \gamma _{2}\gamma
_{3}-\gamma _{4}\gamma _{1}\right) \left( \alpha _{1}\alpha _{4}-\alpha
_{3}\alpha _{2}\right) /3,  \nonumber \\
&&a_{3}a_{5}-a_{1}a_{7}=\allowbreak \gamma _{3}^{2}\left( -\beta _{2}\beta
_{3}+\beta _{1}\beta _{4}\right) \left( \alpha _{1}\alpha _{4}-\alpha
_{3}\alpha _{2}\right) /3,  \nonumber \\
&&a_{2}a_{4}-a_{0}a_{6}=\allowbreak \gamma _{1}^{2}\left( -\beta _{2}\beta
_{3}+\beta _{1}\beta _{4}\right) \left( \alpha _{1}\alpha _{4}-\alpha
_{3}\alpha _{2}\right) /3,  \nonumber \\
&&a_{0}a_{7}-a_{3}a_{4}=\left( \alpha _{1}\alpha _{4}-\alpha _{2}\alpha
_{3}\right) (\gamma _{1}\gamma _{3}(\beta _{2}\beta _{3}-\beta _{1}\beta
_{4})+\beta _{1}\beta _{3}(\gamma _{2}\gamma _{3}-\gamma _{1}\gamma _{4}))/3,
\nonumber \\
&&a_{1}a_{6}-a_{2}a_{5}=-\left( \alpha _{1}\alpha _{4}-\alpha _{2}\alpha
_{3}\right) (\gamma _{1}\gamma _{3}(\beta _{1}\beta _{4}-\beta _{2}\beta
_{3})+\beta _{1}\beta _{3}(\gamma _{2}\gamma _{3}-\gamma _{1}\gamma _{4}))/3,
\nonumber \\
&&a_{0}a_{7}-a_{2}a_{5}=\allowbreak \left( \beta _{1}\beta _{4}-\beta
_{2}\beta _{3}\right) (\alpha _{1}\alpha _{3}(\gamma _{2}\gamma _{3}-\gamma
_{1}\gamma _{4})+\gamma _{1}\gamma _{3}(\alpha _{2}\alpha _{3}-\alpha
_{1}\alpha _{4}))/3,  \nonumber \\
&&a_{1}a_{6}-a_{3}a_{4}\allowbreak =-\allowbreak \left( \beta _{1}\beta
_{4}-\beta _{2}\beta _{3}\right) (\alpha _{1}\alpha _{3}(\gamma _{2}\gamma
_{3}-\gamma _{1}\gamma _{4})+\gamma _{1}\gamma _{3}(\alpha _{1}\alpha
_{4}-\alpha _{2}\alpha _{3}))/3.  \nonumber
\end{eqnarray}%
By using the above equations, we can conclude that $\psi \rangle $ is
equivalent to $|W\rangle $\ under SLOCC if and only if $a_{i}$\ satisfy the
following equation 
\begin{eqnarray}
(a_{0}a_{7}-a_{2}a_{5}+(a_{1}a_{6}-a_{3}a_{4}))^{2}-4(a_{2}a_{4}-a_{0}a_{6})(a_{3}a_{5}-a_{1}a_{7})=0
\label{w2}
\end{eqnarray}
and inequalities 
\begin{eqnarray}
&&a_{0}a_{3}\neq a_{1}a_{2}\vee a_{5}a_{6}\neq a_{4}a_{7},  \nonumber \\
&&a_{1}a_{4}\neq a_{0}a_{5}\vee a_{3}a_{6}\neq a_{2}a_{7},  \nonumber \\
&&a_{3}a_{5}\neq a_{1}a_{7}\vee a_{2}a_{4}\neq a_{0}a_{6}.
\end{eqnarray}%
Notice that from (\ref{ghz4}) Eq. (\ref{w2}) can be replaced by any one of
the following equations.

\begin{eqnarray}
&&(a_{0}a_{7}-a_{3}a_{4}-(a_{1}a_{6}-a_{2}a_{5}))^{2}-4(a_{1}a_{4}-a_{0}a_{5})(a_{3}a_{6}-a_{2}a_{7})=0,
\nonumber \\
&&(a_{0}a_{7}-a_{2}a_{5}-(a_{1}a_{6}-a_{3}a_{4}))^{2}-4(a_{0}a_{3}-a_{1}a_{2})(a_{4}a_{7}-a_{5}a_{6})=0.
\nonumber \\
&&
\end{eqnarray}

\subsection{A-BC class}

If $|\psi \rangle $ belongs to A-BC class, then $|\psi \rangle $ can be
written as $|\psi \rangle =(s|0\rangle _{A}+t|1\rangle _{A})(a|00\rangle
_{BC}+b|01\rangle _{BC}+c|10\rangle _{BC}+d|11\rangle _{BC})$, where $bc\neq
ad$ since systems $B$ and $C$ are entangled. Thus we obtain the following
equations. 
\begin{eqnarray}
&&as=a_{0},~~~~bs=a_{1},~~~~cs=a_{2},~~~~ds=a_{3},  \nonumber \\
&&at=a_{4},~~~~bt=a_{5},~~~~ct=a_{6},~~~~dt=a_{7}.  \nonumber \\
&&  \label{c2}
\end{eqnarray}%
By using the equations (\ref{c2}) and $bc\neq ad$, we can obtain the
following equalities and inequalities, 
\begin{eqnarray}
&&a_{0}a_{5}=a_{1}a_{4},~~a_{2}a_{7}=a_{3}a_{6},  \nonumber \\
&&a_{0}a_{6}=a_{2}a_{4},~~a_{1}a_{7}=a_{3}a_{5},  \nonumber \\
&&a_{0}a_{7}=a_{3}a_{4},~~a_{1}a_{6}=a_{2}a_{5};  \nonumber \\
&&a_{1}a_{2}\neq a_{0}a_{3}\vee a_{5}a_{6}\neq a_{4}a_{7}.  \label{c1}
\end{eqnarray}%
It is clear that the relations in Eq. (\ref{c1}) are the necessary condition
of $|\psi \rangle $ being equivalent to the class A-BC. Conversely, we can
show that this criterion is sufficient too. To this end, let $\left|
s\right| ^{2}=\left| a_{0}\right| ^{2}+\left| a_{1}\right| ^{2}+\left|
a_{2}\right| ^{2}+\left| a_{3}\right| ^{2}$, $\left| t\right| ^{2}=\left|
a_{4}\right| ^{2}+\left| a_{5}\right| ^{2}+\left| a_{6}\right| ^{2}+\left|
a_{7}\right| ^{2}$, $\left| a\right| ^{2}=\left| a_{0}\right| ^{2}+\left|
a_{4}\right| ^{2}$, $\left| b\right| ^{2}=\left| a_{1}\right| ^{2}+\left|
a_{5}\right| ^{2}$, $\left| c\right| ^{2}=\left| a_{2}\right| ^{2}+\left|
a_{6}\right| ^{2}$ and $\left| d\right| ^{2}=\left| a_{3}\right| ^{2}+\left|
a_{7}\right| ^{2}$. For the real number case, we can see that the above
amplitude equations in (\ref{c2}) hold true under the equalities in (\ref{c1}%
). For example, $%
a^{2}s^{2}=(a_{0}^{2}+a_{4}^{2})(a_{0}^{2}+a_{1}^{2}+a_{2}^{2}+a_{3}^{2})=a_{0}^{2}(a_{0}^{2}+a_{1}^{2}+a_{2}^{2}+a_{3}^{2})+a_{0}^{2}a_{4}^{2}+a_{1}^{2}a_{4}^{2}+a_{2}^{2}a_{4}^{2}+a_{3}^{2}a_{4}^{2}=a_{0}^{2}(a_{0}^{2}+a_{1}^{2}+a_{2}^{2}+a_{3}^{2}+a_{4}^{2}+a_{5}^{2}+a_{6}^{2}+a_{7}^{2})=a_{0}^{2} 
$. These amplitude\ equations imply that $|\psi \rangle =(s|0\rangle
_{A}+t|1\rangle _{A})(a|00\rangle _{BC}+b|01\rangle _{BC}+c|10\rangle
_{BC}+d|11\rangle _{BC})$. This case can be extended to the complex case.

Next we show that systems $B$ and $C$ are entangled. From the amplitude
equations (\ref{c2}) we can obtain that $bcs^{2}=a_{1}a_{2}$, $%
ads^{2}=a_{0}a_{3}$, $bct^{2}=a_{5}a_{6}$ and $adt^{2}=a_{4}a_{7}$. Further
from the inequalities of this criterion, it is easy to derive that $%
bcs^{2}\neq ads^{2}$ or $bct^{2}\neq adt^{2}$. Since at least one of $s$ and 
$t$ is not zero, then $bc\neq ad$, which means that systems $B$ and $C$ are
entangled.

We arrive at the conclusion that $|\psi \rangle$ belongs to the A-BC class
if and only if $a_{i}$ satisfy the equalities and inequalities given in Eq. (%
\ref{c1}).

Notice that it can be verified by using (\ref{c2}) that (\ref{w2}) also
holds for $A-BC$ class.

\subsection{B-AC class}

Similarly, we can show that $|\psi \rangle $\ belongs to this class if and
only if $a_{i}$ satisfy the following equalities and inequalities: 
\begin{eqnarray}
&&a_{0}a_{3}=a_{1}a_{2},~~~~a_{4}a_{7}=a_{5}a_{6},  \nonumber \\
&&a_{0}a_{6}=a_{2}a_{4},~~~~a_{1}a_{7}=a_{3}a_{5},  \nonumber \\
&&a_{0}a_{7}=a_{2}a_{5}~~~~a_{1}a_{6}=a_{3}a_{4};  \nonumber \\
&&a_{1}a_{4}\neq a_{0}a_{5} \lor a_{3}a_{6}\neq a_{2}a_{7}.  \label{d1}
\end{eqnarray}

The proof is similar to the one for class $A-BC$.

Notice that (\ref{w2}) also holds for $B-AC$ class.

\subsection{C-AB class}

$|\psi \rangle $\ belongs to the class C-AB, if and only if $a_{i}$ satisfy
the following equalities and inequalities: 
\begin{eqnarray}
&&a_{0}a_{3}=a_{1}a_{2},~~~~a_{4}a_{7}=a_{5}a_{6},  \nonumber \\
&&a_{0}a_{5}=a_{1}a_{4},~~~~a_{2}a_{7}=a_{3}a_{6},  \nonumber \\
&&a_{0}a_{7}=a_{1}a_{6},~~~~a_{2}a_{5}=a_{3}a_{4};  \nonumber \\
&& a_{2}a_{4}\neq a_{0}a_{6} \lor a_{3}a_{5}\neq a_{1}a_{7}.  \label{e1}
\end{eqnarray}

The proof is similar to the one for class $A-BC$.

Notice that (\ref{w2}) also holds for $C-AB$ class.

\subsection{$A-B-C$ class}

If the state $|\psi \rangle $ belongs to the class A-B-C, the necessary and
sufficient condition\cite{LDF} reads 
\begin{eqnarray}
&&a_{0}a_{3}=a_{1}a_{2},~~~~a_{5}a_{6}=a_{4}a_{7},  \nonumber \\
&&a_{0}a_{5}=a_{1}a_{4},~~~~a_{3}a_{6}=a_{2}a_{7},  \nonumber \\
&&a_{1}a_{7}=a_{3}a_{5},~~~~a_{2}a_{4}=a_{0}a_{6},  \nonumber \\
&&a_{0}a_{7}=a_{1}a_{6},~~~~a_{2}a_{5}=a_{3}a_{4}.  \label{f1}
\end{eqnarray}

Notice that it is easy to see from (\ref{f1}) that (\ref{w2}) holds.

\subsection{The complete partition}

Before proceeding further, we would like to point out that the criteria for
classes $A-B-C$, $A-BC$, $B-AC$, $C-AB$, $|GHZ\rangle $ and $|W\rangle $ are
exclusive to each other ( see Appendix A for the details). The criteria form
a complete partition (see table 1 for the details). For this completeness,
what we need to do next is to demonstrate the ``not-occur'' cases and the
alternate criteria for $A-BC$, $B-AC$, $C-AB$, $A-B-C$ in table 1.\ We
finish these proofs in appendixes $B$, $C$ and $D$ respectively. Thus we
argue that any state has to be in one of classes $A-B-C$, $A-BC$, $B-AC$, $%
C-AB$, $|GHZ\rangle $ and $|W\rangle $ and any two states satisfying the
same criterion are related by SLOCC. In other words, we give a different
proof of D\"{u}r et al.'s SLOCC classification.

The table 1: The complete partition

\hoffset-2cm 
\begin{tabular}{|c|c|c|c|r|}
\hline
$(a_{0}a_{7}-a_{2}a_{5}+(a_{1}a_{6}-a_{3}a_{4}))^{2}$ & $%
a_{0}a_{3}=a_{1}a_{2}$ & $a_{1}a_{4}=a_{0}a_{5}$ & $a_{3}a_{5}=a_{1}a_{7}$ & 
\\ 
$=4(a_{2}a_{4}-a_{0}a_{6})(a_{3}a_{5}-a_{1}a_{7})$ & $\wedge
a_{5}a_{6}=a_{4}a_{7}$ & $\wedge a_{3}a_{6}=a_{2}a_{7}$ & $\wedge
a_{2}a_{4}=a_{0}a_{6}$ &  \\ \hline
N & \multicolumn{4}{|r|}{$|GHZ\rangle $} \\ \hline
&  & N & N & $|W\rangle $ \\ \cline{4-5}
& N &  & Y & not-occur \\ \cline{3-5}
&  & Y & N & not-occur \\ \cline{4-5}
Y &  &  & Y & $A-BC$ \\ \cline{2-5}
&  & N & N & not-occur \\ \cline{4-5}
& Y &  & Y & $B-AC$ \\ \cline{3-5}
&  & Y & N & $C-AB$ \\ \cline{4-5}
&  &  & Y & $A-B-C$ \\ \hline
\end{tabular}

In the table 1, ``Y'' means that the condition holds and ``N'' means that
the condition does not hold. ``not-occur'' means the case does not occur. \ 

\section{Classification of entanglement for a four-qubit system}

We now turn the discussion to four-qubit systems. By means of the criteria
for SLOCC entanglement classes of three-qubits, we can derive the criteria
for degenerated four-qubit entanglement. We give the necessary criteria
which the four-qubit $|GHZ\rangle $ and $|W\rangle $\ classes satisfy. Let $%
|C_{4}\rangle =(|0011\rangle +|0110\rangle +|1100\rangle +|1010\rangle
+|1001\rangle +|0101\rangle )/\sqrt{6}$. We argue that $|C_{4}\rangle $ is a
genuinely four-qubit entangled state which is inequivalent to $|GHZ\rangle $
or $|W\rangle $ states of four-qubits under SLOCC.

Let $|\psi \rangle $ $=\sum_{j=0}^{15}a_{j}|j\rangle $ be any pure state of
four-qubits.

\subsection{Three-qubit GHZ entanglement accompanied with a separable one
qubit}

We only study that $ABC$ are $GHZ-$ entangled. Let $|\psi \rangle =|\varphi
\rangle _{ABC}(s|0\rangle +t|1\rangle )_{D}$, where$|\varphi \rangle
=\sum_{i=0}^{7}b_{i}|i\rangle $ and $|\varphi \rangle $ is in $|GHZ\rangle $
class of three-qubits. By the criterion for $|GHZ\rangle $ class of
three-qubits, we have the following inequality,

$(b_{0}b_{7}-b_{2}b_{5}+(b_{1}b_{6}-b_{3}b_{4}))^{2}\neq
4(b_{2}b_{4}-b_{0}b_{6})(b_{3}b_{5}-b_{1}b_{7})$.

Since $\gamma \neq 0$ or $\delta \neq 0$,

$%
[(b_{0}b_{7}-b_{2}b_{5}+(b_{1}b_{6}-b_{3}b_{4}))^{2}-4(b_{2}b_{4}-b_{0}b_{6})(b_{3}b_{5}-b_{1}b_{7})]\gamma ^{4}\neq 0 
$ or

$%
[(b_{0}b_{7}-b_{2}b_{5}+(b_{1}b_{6}-b_{3}b_{4}))^{2}-4(b_{2}b_{4}-b_{0}b_{6})(b_{3}b_{5}-b_{1}b_{7})]\delta ^{4}\neq 0 
$.

Consequently, $a_{i}$ satisfy the following inequalities:

$(a_{0}a_{14}-a_{4}a_{10}+a_{2}a_{12}-a_{6}a_{8})^{2}\neq
4(a_{4}a_{8}-a_{0}a_{12})(a_{6}a_{10}-a_{2}a_{14})$ or

$(a_{1}a_{15}-a_{5}a_{11}+a_{3}a_{13}-a_{7}a_{9})^{2}\neq
4(a_{5}a_{9}-a_{1}a_{13})(a_{7}a_{11}-a_{3}a_{15})$

and the following equalities:

$a_{i}a_{j}=a_{k}a_{l}$, where $a_{i}a_{j}-a_{k}a_{l}$ are all the $2\times
2 $ minor determinants of the following matrix,

$\left( 
\begin{tabular}{cccccccc}
$a_{0}$ & $a_{2}$ & $a_{4}$ & $a_{6}$ & $a_{8}$ & $a_{10}$ & $a_{12}$ & $%
a_{14}$ \\ 
$a_{1}$ & $a_{3}$ & $a_{5}$ & $a_{7}$ & $a_{9}$ & $a_{11}$ & $a_{13}$ & $%
a_{15}$%
\end{tabular}%
\right) $.

These conditions are necessary and sufficient.

Example, $(|0000\rangle +|1110\rangle )/\sqrt{2}$ satisfies the above
conditions.

\subsection{Three-qubit W entanglement accompanied with a separable one qubit%
}

We only illustrate that $ABC$ are $W-$ entangled. Let $|\psi \rangle
=|\varphi \rangle _{ABC}(s|0\rangle +t|1\rangle )_{D}$, where$|\varphi
\rangle =\sum_{i=0}^{7}b_{i}|i\rangle $ and $|\varphi \rangle $\ is in $%
|W\rangle $ class of three-qubits. By means of the criterion for $|W\rangle $
class of three-qubits, $a_{i}$ satisfy the following equalities:

$%
(a_{0}a_{14}-a_{4}a_{10}+a_{2}a_{12}-a_{6}a_{8})^{2}=4(a_{4}a_{8}-a_{0}a_{12})(a_{6}a_{10}-a_{2}a_{14}) 
$,

$%
(a_{1}a_{15}-a_{5}a_{11}+a_{3}a_{13}-a_{7}a_{9})^{2}=4(a_{5}a_{9}-a_{1}a_{13})(a_{7}a_{11}-a_{3}a_{15}), 
$

$a_{i}a_{j}=a_{k}a_{l}$, where $a_{i}a_{j}-a_{k}a_{l}$ are all the $2\times
2 $ minor determinants of the following matrix,

$\left( 
\begin{tabular}{cccccccc}
$a_{0}$ & $a_{2}$ & $a_{4}$ & $a_{6}$ & $a_{8}$ & $a_{10}$ & $a_{12}$ & $%
a_{14}$ \\ 
$a_{1}$ & $a_{3}$ & $a_{5}$ & $a_{7}$ & $a_{9}$ & $a_{11}$ & $a_{13}$ & $%
a_{15}$%
\end{tabular}%
\right) $

and satisfy the following inequalities:

$(a_{3}a_{5}\neq a_{1}a_{7}\vee $ $a_{10}a_{12}\neq a_{8}a_{14}\vee $ $%
a_{2}a_{4}\neq a_{0}a_{6}\vee a_{11}a_{13}\neq a_{9}a_{15})\wedge $

$(a_{2}a_{8}\neq a_{0}a_{10}\vee $ $a_{3}a_{9}\neq a_{1}a_{11}\vee $ $%
a_{6}a_{12}\neq a_{4}a_{14}\vee a_{7}a_{13}\neq a_{5}a_{15})\wedge $

$(a_{6}a_{10}\neq a_{2}a_{14}\vee $ $a_{7}a_{11}\neq a_{3}a_{15}\vee $ $%
a_{4}a_{8}\neq a_{0}a_{12}\vee a_{5}a_{9}\neq a_{1}a_{13})$.

These conditions are necessary and sufficient.

For example, $|W\rangle _{123}\otimes |0\rangle _{4}$ and $(|110\rangle
_{123}+|101\rangle _{123}+|011\rangle _{123})\otimes |0\rangle _{4}$ satisfy
the above conditions.

\subsection{ A state consisting of two $EPR$ pairs}

We only investigate $AB-CD$ class, where $AB$ and $CD$ are $EPR$ pairs, as
follows. $|\psi \rangle $ is in this class if and only if $a_{i}$ satisfy
the following inequalities

$(a_{4}a_{8}\neq a_{0}a_{12}\vee a_{6}a_{10}\neq a_{2}a_{14}\vee
a_{5}a_{9}\neq a_{1}a_{13}\vee a_{7}a_{11}\neq a_{3}a_{15})\wedge $

$(a_{1}a_{2}\neq a_{0}a_{3}\vee a_{5}a_{6}\neq a_{4}a_{7}\vee
a_{9}a_{10}\neq a_{8}a_{11}\vee a_{13}a_{14}\neq a_{12}a_{15})$

and the following equalities:

$a_{i}a_{j}=a_{k}a_{l}$, where $a_{i}a_{j}-a_{k}a_{l}$ are all the $2\times
2 $ minor determinants of the following matrix,

$\left( 
\begin{tabular}{cccc}
$a_{0}$ & $a_{1}$ & $a_{2}$ & $a_{3}$ \\ 
$a_{4}$ & $a_{5}$ & $a_{6}$ & $a_{7}$ \\ 
$a_{8}$ & $a_{9}$ & $a_{10}$ & $a_{11}$ \\ 
$a_{12}$ & $a_{13}$ & $a_{14}$ & $a_{15}$%
\end{tabular}%
\right) $.

Example, $|\phi _{4}\rangle $, which is $(|0000\rangle +|0011\rangle
+|1100\rangle -|1111\rangle )/2$ in \cite{Briegel}, does not satisfy the
above conditions.

\subsection{Only two qubits are entangled}

We only discuss $A-B-CD$ class where $CD$ is a $EPR$ pair. Then one can
obtain that $|\psi \rangle $ is in this class if and only if $a_{i}$ satisfy
the following inequalities

$a_{1}a_{2}\neq a_{0}a_{3}\vee a_{5}a_{6}\neq a_{4}a_{7}\vee a_{9}a_{10}\neq
a_{8}a_{11}\vee a_{13}a_{14}\neq a_{12}a_{15}$

and the following equalities

$a_{i}a_{j}=a_{k}a_{l}$, where $i+j=k+l$ and $i\oplus j=k\oplus l$, $i<j$, $%
k<l$, $i=l$ $(mod 4)$, $j=k$ $(mod 4)$.

\subsection{A-B-C-D class}

$|\psi \rangle $ is separable if and only if $a_{i}a_{j}=a_{k}a_{l}$, where $%
i+j=k+l$ and $i\oplus j=k\oplus l$.

\subsection{$|GHZ\rangle $ class}

Let $|GHZ\rangle =(|0000\rangle +|1111\rangle )/\sqrt{2}$. Then if $|\psi
\rangle $\ is equivalent to $|GHZ\rangle $\ under SLOCC then $a_{i}$\
satisfy the following inequality,

$a_{2}a_{13}-a_{3}a_{12}+(a_{4}a_{11}-a_{5}a_{10})\neq
(a_{0}a_{15}-a_{1}a_{14})+(a_{6}a_{9}-a_{7}a_{8})$

and the following equalities,

$%
(a_{1}a_{4}-a_{0}a_{5})(a_{11}a_{14}-a_{10}a_{15})=(a_{3}a_{6}-a_{2}a_{7})(a_{9}a_{12}-a_{8}a_{13}) 
$,

$%
(a_{4}a_{7}-a_{5}a_{6})(a_{8}a_{11}-a_{9}a_{10})=(a_{0}a_{3}-a_{1}a_{2})(a_{12}a_{15}-a_{13}a_{14}), 
$

$\allowbreak
(a_{3}a_{5}-a_{1}a_{7})(a_{10}a_{12}-a_{8}a_{14})=(a_{2}a_{4}-a_{0}a_{6})(a_{11}a_{13}-a_{9}a_{15}) 
$.

For example, $|GHZ\rangle $ satisfies the above conditions, while $|\phi
_{4}\rangle $\cite{Briegel} does not satisfy the second equality of this
criterion. Thus, we also show that $|\phi _{4}\rangle $ is not in the
four-qubit $GHZ$ entanglement class.

\subsection{$|W\rangle $ class}

Let $|W\rangle =$\ $(|0001\rangle +|0010\rangle +|0100\rangle +|1000)/2$.
Then if $|\psi \rangle $\ is equivalent to $|W\rangle $\ under SLOCC then $%
a_{i}$\ satisfy the following equalities

$a_{2}a_{13}-a_{3}a_{12}+a_{4}a_{11}-$ $a_{5}a_{10}=a_{0}a_{15}-$ $%
a_{1}a_{14}+a_{6}a_{9}-a_{7}a_{8}$

$%
((a_{0}a_{7}-a_{3}a_{4})+(a_{1}a_{6}-a_{2}a_{5}))^{2}=4(a_{3}a_{5}-a_{1}a_{7})(a_{2}a_{4}-a_{0}a_{6}), 
$

$%
((a_{4}a_{11}-a_{7}a_{8})+(a_{5}a_{10}-a_{6}a_{9}))^{2}=4(a_{7}a_{9}-a_{5}a_{11})(a_{6}a_{8}-a_{4}a_{10}) 
$,

$%
((a_{8}a_{15}-a_{11}a_{12})+(a_{9}a_{14}-a_{10}a_{13}))^{2}=4(a_{11}a_{13}-a_{9}a_{15})(a_{10}a_{12}-a_{8}a_{14}) 
$,

$%
(a_{0}a_{14}-a_{4}a_{10}+a_{2}a_{12}-a_{6}a_{8})^{2}=4(a_{4}a_{8}-a_{0}a_{12})(a_{6}a_{10}-a_{2}a_{14}) 
$

$%
(a_{1}a_{15}-a_{5}a_{11}+a_{3}a_{13}-a_{7}a_{9})^{2}=4(a_{5}a_{9}-a_{1}a_{13})(a_{7}a_{11}-a_{3}a_{15}) 
$

$%
(a_{0}a_{11}-a_{2}a_{9}+a_{1}a_{10}-a_{3}a_{8})^{2}=4(a_{2}a_{8}-a_{0}a_{10})(a_{3}a_{9}-a_{1}a_{11}) 
$

$%
(a_{4}a_{15}-a_{6}a_{13}+a_{5}a_{14}-a_{7}a_{12})^{2}=4(a_{6}a_{12}-a_{4}a_{14})(a_{7}a_{13}-a_{5}a_{15}) 
$

$%
(a_{0}a_{13}-a_{4}a_{9}+a_{1}a_{12}-a_{5}a_{8})^{2}=4(a_{4}a_{8}-a_{0}a_{12})(a_{5}a_{9}-a_{1}a_{13}) 
$

$%
(a_{2}a_{15}-a_{6}a_{11}+a_{3}a_{14}-a_{7}a_{10})^{2}=4(a_{6}a_{10}-a_{2}a_{14})(a_{7}a_{11}-a_{3}a_{15}) 
$

$%
(a_{1}a_{4}-a_{0}a_{5})(a_{11}a_{14}-a_{10}a_{15})=(a_{3}a_{6}-a_{2}a_{7})(a_{9}a_{12}-a_{8}a_{13}), 
$

$%
(a_{4}a_{7}-a_{5}a_{6})(a_{8}a_{11}-a_{9}a_{10})=(a_{0}a_{3}-a_{1}a_{2})(a_{12}a_{15}-a_{13}a_{14}), 
$

$%
(a_{3}a_{5}-a_{1}a_{7})(a_{10}a_{12}-a_{8}a_{14})=(a_{2}a_{4}-a_{0}a_{6})(a_{11}a_{13}-a_{9}a_{15}), 
$

and the following inequalities

$a_{0}a_{3}\neq $\ $a_{1}a_{2}$\ or $a_{5}a_{6}\neq $\ $a_{4}a_{7}$\ or $%
a_{8}a_{11}\neq $\ $a_{9}a_{10}$\ or $a_{13}a_{14}\neq $\ $a_{12}a_{15}$,

$a_{1}a_{4}\neq $\ $a_{0}a_{5}$\ or $a_{3}a_{6}\neq a_{2}a_{7}$\ or $%
a_{9}a_{12}\neq $\ $a_{8}a_{13}$\ or $a_{11}a_{14}\neq $\ $a_{10}a_{15}$,

and $a_{3}a_{5}\neq a_{1}a_{7}$\ or $a_{2}a_{4}\neq $\ $a_{0}a_{6}$\ or $%
a_{11}a_{13}\neq $\ $a_{9}a_{15}$\ or $a_{10}a_{12}\neq $\ $a_{8}a_{14}$.

The proof is in appendix $E$.

Example, $|W\rangle $ satisfies the above conditions, while $|\phi
_{4}\rangle $ does not satisfy this criterion.

\subsection{A genuinely four-qubit entanglement $|C_{4}\rangle $ class}

It is easy to verify that $|C_{4}\rangle $ does not satisfy the criteria for
degenerated four-qubit entanglement. It means that $|C_{4}\rangle $ is a
genuinely four-qubit entangled state. We can also observe that $%
|C_{4}\rangle $ does not satisfy the first equality of the criteria for $%
|GHZ\rangle $ or $|W\rangle $ classes. Therefore, $|C_{4}\rangle $ is
inequivalent to $|GHZ\rangle $ or $|W\rangle $ under SLOCC.

$|C_{4}\rangle $ possesses the following properties.

(1).$|C_{4}\rangle $ is symmetric under permutation of parties.

(2).$|C_{4}\rangle $ is self-complementary in the following sense.

Let $\bar{1}$ ( $\bar{0}$ ) be the complement of a bit 1 $(0)$. Then $\bar{0}
$ $=1$ and $\bar{1}=0$. Let $\bar{z}=\bar{z_{1}}\bar{z_{2}}...\bar{z_{n}}$
denote the complement of a binary string $z=z_{1}z_{2}....z_{n}$. Likewise,
we can define $|\overline{\Phi }\rangle =c_{0}|\bar{0}\rangle $ $+c_{1}|\bar{%
1}\rangle $ $+....+c_{2^{n}-1}|\overline{(2^{n}-1)}\rangle $, where $|\Phi
\rangle =c_{0}|0\rangle $ $+c_{1}|1\rangle $ $+....+c_{2^{n}-1}|2^{n}-1%
\rangle $, and call $|\overline{\Phi }\rangle $ the complement of $|\Phi
\rangle $. Obviously, $|C_{4}\rangle =|\overline{C_{4}}\rangle $.

(3). When any one of the four qubits is traced out, the remaining three
qubits are identical and mixed. For example, $tr_{D}(|C_{4}\rangle \langle
C_{4}|)=(|W\rangle \langle W|+|\overline{W}\rangle \langle \overline{W}|)/2$%
, where $|W\rangle $ is $(|001\rangle +|010\rangle +|100\rangle )/\sqrt{3}$.

(4). When any two of the four qubits are traced out, the remaining two
qubits are identical and mixed. For example, $tr_{CD}(|C_{4}\rangle \langle
C_{4}|)=\frac{1}{6}(|11\rangle \langle 11|+|00\rangle \langle 00|)+\frac{2}{3%
}|\Psi ^{+}\rangle \langle \Psi ^{+}|$, where $|\Psi ^{+}\rangle
=(|01\rangle +|10\rangle )/\sqrt{2}$.

\section{Summaries}

In summaries, in this paper we propose the simple criteria, in which only
addition and multiplication occur, for the SLOCC\ equivalence classes and
show that these criteria are exclusive and form a complete partition. Thus,
new proofs are given for D\"{u}r et al.'s SLOCC classification of
three-qubits. By means of the criteria for SLOCC entanglement classes of
three-qubits, we can derive the criteria for degenerated four-qubit
entanglement. We obtain the necessary criteria which the four-qubit $%
|GHZ\rangle $ and $|W\rangle $\ classes satisfy. We observe that $%
|C_{4}\rangle $ is a genuinely four-qubit entangled state which is
inequivalent to $|GHZ\rangle $ or $|W\rangle $ states of four-qubits under
SLOCC. By means of the criteria of four-qubits, we can determine if a state
is a genuinely four-qubit entangled state which is inequivalent to $%
|GHZ\rangle $ or $|W\rangle $ under SLOCC.

Appendix A

Proof.

Case 1. The criteria for $|GHZ\rangle $ and $|W\rangle $ classes are
exclusive.

This is because the criterion for $|GHZ\rangle $ class contradicts condition
(1) of the criterion for $|W\rangle $ class.

Case 2.$\ $The criteria for class $|W\rangle $ and for any one of classes $%
A-B-C$ , $A-BC$, $B-AC$ and $C-AB$ are exclusive.

Condition (2) of the criterion\ for $|W\rangle $ class implies that the
criteria for $|W\rangle $ class and $A-B-C$ class are exclusive.

In $A-BC$ class, any state satisfies $a_{1}a_{4}=a_{0}a_{5}$ and $%
a_{3}a_{6}=a_{2}a_{7}$. However, condition (2) of the criterion for $%
|W\rangle $ class says that $a_{1}a_{4}\neq $ $a_{0}a_{5}$ or $%
a_{3}a_{6}\neq a_{2}a_{7}$. Therefore the criteria for $|W\rangle $ class
and $A-BC$ class are exclusive.

Similarly, the criteria for classes $|W\rangle $ and $B-AC$ and for classes $%
|W\rangle $ and $C-AB$ are exclusive, respectively.

Case 3. The criteria for class $|GHZ\rangle $ and for any one of classes $%
A-B-C$ , $A-BC$, $B-AC$ and $C-AB$ are exclusive.

First let us demonstrate that the criteria for classes $|GHZ\rangle $ and $%
A-B-C$ are exclusive. We see immediately that any state in $A-B-C$ class
does not satisfy the criteria for $|GHZ\rangle $ class. Conversely, if a
state in $|GHZ\rangle $ class satisfies $a_{3}a_{5}\neq a_{1}a_{7}$ or $%
a_{2}a_{4}\neq $ $a_{0}a_{6}$, then the state is not separable by the
criterion for class $A-B-C$. Otherwise the state in $|GHZ\rangle $ class
satisfies$\ a_{3}a_{5}=a_{1}a_{7}$ and $a_{2}a_{4}=$ $a_{0}a_{6}$. By the
criterion for $|GHZ\rangle $ class, it follows that $a_{0}a_{7}-$ $%
a_{2}a_{5}+(a_{1}a_{6}-a_{3}a_{4})\neq 0$. It means that it is impossible
that $a_{0}a_{7}=a_{2}a_{5}$ and $a_{1}a_{6}=a_{3}a_{4}$. Thus the state in $%
|GHZ\rangle $ class is not in class $A-B-C$. Therefore the criteria for
classes $|GHZ\rangle $ and $A-B-C$ are exclusive. \ \ \ \ \ \ \ \ 

Next let us deduce that the criteria for classes $|GHZ\rangle $ and $A-BC$
are exclusive. From section 2 we know that any state in $A-BC$ class
satisfies $a_{1}a_{4}=a_{0}a_{5}$, $a_{3}a_{6}=a_{2}a_{7}$, $a_{0}a_{7}=$ $%
a_{3}a_{4}$ and $a_{1}a_{6}=a_{2}a_{5}$. Using these equalities we can
derive that $a_{0}a_{7}-a_{2}a_{5}+(a_{1}a_{6}-a_{3}a_{4})=0$ and $%
(a_{2}a_{4}-a_{0}a_{6})(a_{3}a_{5}-a_{1}a_{7})=0$. It means that any state
in $A-BC$ class does not satisfy criterion for $|GHZ\rangle $ class.
Conversely, if a state in $|GHZ\rangle $ class satisfies $a_{0}a_{7}\neq $ $%
a_{3}a_{4}$ or $a_{1}a_{6}\neq a_{2}a_{5}$, then the state in $|GHZ\rangle $
class is not in $A-BC$ class by the criterion for class $A-BC$. Otherwise
the state in $|GHZ\rangle $ class satisfies $a_{0}a_{7}=$ $a_{3}a_{4}$ and $%
a_{1}a_{6}=a_{2}a_{5}$. Then by the criterion for $|GHZ\rangle $ class, it
follows that $(a_{2}a_{4}-a_{0}a_{6})(a_{3}a_{5}-a_{1}a_{7})\neq 0$. This
results in that the state in $|GHZ\rangle $ class\ is not in $A-BC$ class.
Therefore the criteria for classes $|GHZ\rangle $ and $A-BC$ are exclusive.

Similarly, we can infer that the criteria for classes $|GHZ\rangle $ and $%
B-AC$, and for classes $|GHZ\rangle $ and $C-AB$ are exclusive, respectively.

Case 4. It is easy to see that the criteria for classes $A-B-C$ , $A-BC$, $%
B-AC$ and $C-AB$ are pairwise exclusive.

Appendix B

Lemma. No states satisfy the following conditions. In other words, the
conditions (1), (2), (3), (4) and (5) are inconsistent. Therefore the
situation is not applicable. For other ``not-occur'' cases in table 1, the
discussions are similar therefore omitted.

$%
(a_{0}a_{7}-a_{2}a_{5}+(a_{1}a_{6}-a_{3}a_{4}))^{2}=4(a_{2}a_{4}-a_{0}a_{6})(a_{3}a_{5}-a_{1}a_{7}).....(1) 
$

$a_{3}a_{5}=a_{1}a_{7}$......(2)

$a_{2}a_{4}=a_{0}a_{6}$ .......(3)

$a_{0}a_{3}\neq $\ $a_{1}a_{2}$\ or $a_{5}a_{6}\neq $\ $a_{4}a_{7}$ ......(4)

$a_{1}a_{4}\neq $\ $a_{0}a_{5}$\ or $a_{3}a_{6}\neq a_{2}a_{7}$........(5)

Proof.

From (1), (2) and (3), we have the following

$a_{0}a_{7}-a_{3}a_{4}=a_{2}a_{5}-a_{1}a_{6}$...........(6).

Next we prove that (4) holds means that (5) does not hold under (1), (2) and
(3). That is, $a_{0}a_{3}\neq $$a_{1}a_{2}$ or $a_{5}a_{6}\neq $\textbf{\ }$%
a_{4}a_{7}$ results in $a_{1}a_{4}=$ $a_{0}a_{5}$\textbf{\ }and $%
a_{3}a_{6}=a_{2}a_{7}$. There are two cases.

Case 1. $a_{0}a_{3}\neq $\textbf{\ }$a_{1}a_{2}$ implies $a_{1}a_{4}=$%
\textbf{\ }$a_{0}a_{5}$\textbf{\ }and\textbf{\ }$a_{3}a_{6}=a_{2}a_{7}$.

Case 1.1. Replacing $a_{1}a_{7}$ by $a_{3}a_{5}$ and $a_{0}a_{6}$ by $%
a_{2}a_{4}$ in $%
a_{0}a_{1}(a_{0}a_{7}-a_{3}a_{4})=a_{0}a_{1}(a_{2}a_{5}-a_{1}a_{6})$ from
(6) respectively, and by factoring we have $(a_{0}a_{3}-\
a_{1}a_{2})(a_{1}a_{4}-\ a_{0}a_{5})=0$. Therefore $a_{0}a_{3}\neq $\textbf{%
\ }$a_{1}a_{2}$ yields $a_{1}a_{4}=\ a_{0}a_{5}$.

Case 1.2. Replacing $a_{3}a_{5}$ by $a_{1}a_{7}$\ and $a_{2}a_{4}$ by $%
a_{0}a_{6}$ in $%
a_{2}a_{3}(a_{0}a_{7}-a_{3}a_{4})=a_{2}a_{3}(a_{2}a_{5}-a_{1}a_{6})$ from
(6) respectively, and by factoring we have

$(a_{0}a_{3}-\ a_{1}a_{2})(a_{2}a_{7}-a_{3}a_{6})=0$, which yields $%
a_{3}a_{6}=a_{2}a_{7}$ since $a_{0}a_{3}\neq $\textbf{\ }$a_{1}a_{2}.$

Case 2. Suppose $a_{5}a_{6}\neq $\textbf{\ }$a_{4}a_{7}$. Similarly we can
derive $a_{1}a_{4}=$\textbf{\ }$a_{0}a_{5}$\textbf{\ }and\textbf{\ }$%
a_{3}a_{6}=a_{2}a_{7}$.

Appendix C

Lemma. the criterion for $B-AC$ class is equivalent to the following six
conditions.

$a_{0}a_{7}-a_{3}a_{4}=a_{2}a_{5}-a_{1}a_{6}.....(1)$,

$a_{2}a_{4}=a_{0}a_{6}$ .......(2),

$a_{3}a_{5}=a_{1}a_{7}$\textbf{\ }......(3),

$a_{0}a_{3}=$\ $a_{1}a_{2}$\ ........(4),

$a_{5}a_{6}=$\ $a_{4}a_{7}$ ......(5),

$a_{1}a_{4}\neq $\ $a_{0}a_{5}$\ or $a_{3}a_{6}\neq a_{2}a_{7}$........(6).

For $A-BC$ and $C-AB$, the discussion is similar to this one.

Proof. It is trivial to verify that the criterion for $B-AC$ in section 2
satisfies the above conditions. Conversely, we can prove that the above
conditions satisfy the criterion for $B-AC$. It is enough to show $%
a_{0}a_{7}=$\ $a_{2}a_{5}$ and $a_{1}a_{6}=a_{3}a_{4}$. \noindent Replacing $%
a_{2}a_{4}$ by $a_{0}a_{6}$ and $a_{1}a_{2}$ by $a_{0}a_{3}$ respectively,
in $a_{2}(a_{0}a_{7}-a_{3}a_{4})=a_{2}(a_{2}a_{5}-a_{1}a_{6})$ from (1), we
obtain $a_{0}a_{2}a_{7}=$\ $a_{2}^{2}a_{5}$. When $a_{2}\neq 0$, we have $%
a_{0}a_{7}=$\ $a_{2}a_{5}$. Otherwise it is straightforward to obtain $%
a_{0}a_{7}=$\ $a_{2}a_{5}$ since it is trivial when $a_{0}=0$ and $a_{0}\neq
0$ leads to $a_{3}=a_{6}=0$ from (2) and (4) and further, $a_{0}a_{7}=0$
from (1). Similarly, replacing $a_{1}a_{7}$ by $a_{3}a_{5}$ \ and $%
a_{1}a_{2} $ by $a_{0}a_{3}$ respectively,\ in $%
a_{1}(a_{0}a_{7}-a_{3}a_{4})=a_{1}(a_{2}a_{5}-a_{1}a_{6})$ from (1), we get $%
a_{1}a_{3}a_{4}=a_{1}^{2}a_{6}$. Then we observe $a_{1}a_{6}=a_{3}a_{4}$
when $a_{1}\neq 0$. Otherwise it is easy to find $a_{1}a_{6}=a_{3}a_{4}$.

Appendix D

Lemma. The criterion for $A-B-C$ in section 2 is equivalent to the following
equalities.

$a_{0}a_{7}-a_{3}a_{4}=a_{2}a_{5}-a_{1}a_{6}$, $a_{2}a_{4}=a_{0}a_{6}$, $%
a_{3}a_{5}=a_{1}a_{7}$,

$a_{0}a_{3}=$\ $a_{1}a_{2}$,\ $a_{5}a_{6}=$\ $a_{4}a_{7}$, $a_{1}a_{4}=$\ $%
a_{0}a_{5}$,\ $a_{3}a_{6}=a_{2}a_{7}$.

Proof. It is easy to verify that the criterion for $A-B-C$ satisfies the
above equalities. Conversely, in appendix C, by using the first five
equalities we derive $a_{0}a_{7}=a_{2}a_{5}$\textbf{\ }and\textbf{\ }$%
a_{1}a_{6}=a_{3}a_{4}$. Similarly, we can obtain $a_{0}a_{7}=a_{1}a_{6}$%
\textbf{, }$a_{1}a_{6}=a_{2}a_{5}$\textbf{, }$a_{2}a_{5}=a_{3}a_{4}$, $%
a_{0}a_{7}=a_{3}a_{4}$. Therefore the criterion for $A-B-C$ is satisfied.

Appendix E

The necessary criterion of the entanglement class $|W\rangle $ for a
four-qubit system

Proof.

Let $\alpha $, $\beta $, $\gamma $ and $\delta $\ be operators and $|\psi
\rangle =\sum_{i=0}^{15}a_{i}|i\rangle =$ $\alpha \otimes \beta \otimes
\gamma \otimes \delta |W\rangle $, where $\delta =$ $\left( 
\begin{tabular}{cc}
$\ \delta _{1}$ & $\ \delta _{2}$ \\ 
$\ \delta _{3}$ & $\ \delta _{4}$%
\end{tabular}%
\right) $. Then

$a_{0}=(\alpha _{1}\beta _{1}\gamma _{1}\delta _{2}+\alpha _{1}\beta
_{1}\gamma _{2}\delta _{1}+\alpha _{1}\beta _{2}\gamma _{1}\delta
_{1}+\alpha _{2}\beta _{1}\gamma _{1}\delta _{1})/2$ and other $a_{i}$ are
omitted.

Computing $a_{i}a_{j}-a_{k}a_{l}$, where $i+j=k+l$ and $i\oplus j=k\oplus l$%
, we obtain the following equations:

$a_{0}a_{3}-$ $a_{1}a_{2}=\allowbreak \frac{1}{4}\alpha _{1}^{2}\beta
_{1}^{2}\left( \delta _{1}\delta _{4}-\delta _{3}\delta _{2}\right) \left(
-\gamma _{4}\gamma _{1}+\gamma _{2}\gamma _{3}\right) \allowbreak ,$

$\allowbreak a_{4}a_{7}-a_{5}a_{6}=\allowbreak \frac{1}{4}\alpha
_{1}^{2}\beta _{3}^{2}\left( \delta _{1}\delta _{4}-\delta _{3}\delta
_{2}\right) \left( -\gamma _{4}\gamma _{1}+\gamma _{2}\gamma _{3}\right)
\allowbreak ,$

$a_{1}a_{4}-$ $a_{0}a_{5}=\allowbreak \frac{1}{4}\alpha _{1}^{2}\gamma
_{1}^{2}\left( \delta _{1}\delta _{4}-\delta _{3}\delta _{2}\right) \left(
\beta _{1}\beta _{4}-\beta _{3}\beta _{2}\right) ,$

$a_{3}a_{6}-$ $a_{2}a_{7}=\allowbreak \frac{1}{4}\alpha _{1}^{2}\gamma
_{3}^{2}\left( \delta _{1}\delta _{4}-\delta _{3}\delta _{2}\right) \left(
\beta _{1}\beta _{4}-\beta _{3}\beta _{2}\right) ,$

$a_{3}a_{5}-$ $a_{1}a_{7}=\allowbreak -\frac{1}{4}\alpha _{1}^{2}\delta
_{3}^{2}\left( \beta _{1}\beta _{4}-\beta _{3}\beta _{2}\right) \left(
-\gamma _{4}\gamma _{1}+\gamma _{2}\gamma _{3}\right) \allowbreak ,$

$a_{2}a_{4}-$ $a_{0}a_{6}=\allowbreak -\frac{1}{4}\alpha _{1}^{2}\delta
_{1}^{2}\left( \beta _{1}\beta _{4}-\beta _{3}\beta _{2}\right) \left(
-\gamma _{4}\gamma _{1}+\gamma _{2}\gamma _{3}\right) \allowbreak ,$

$a_{8}a_{11}-$ $a_{9}a_{10}=\allowbreak -\frac{1}{4}\alpha _{3}^{2}\beta
_{1}^{2}\left( -\delta _{3}\delta _{2}+\delta _{1}\delta _{4}\right) \left(
-\gamma _{2}\gamma _{3}+\gamma _{4}\gamma _{1}\right) \allowbreak ,$

$a_{12}a_{15}-a_{13}a_{14}=\allowbreak \frac{1}{4}\alpha _{3}^{2}\beta
_{3}^{2}\left( \delta _{1}\delta _{4}-\delta _{3}\delta _{2}\right) \left(
-\gamma _{4}\gamma _{1}+\gamma _{2}\gamma _{3}\right) ,$

$a_{9}a_{12}-$ $a_{8}a_{13}=\allowbreak \frac{1}{4}\alpha _{3}^{2}\gamma
_{1}^{2}\left( \delta _{1}\delta _{4}-\delta _{3}\delta _{2}\right) \left(
\beta _{1}\beta _{4}-\beta _{3}\beta _{2}\right) ,$

$a_{11}a_{14}-$ $a_{10}a_{15}=\allowbreak \frac{1}{4}\alpha _{3}^{2}\gamma
_{3}^{2}\left( \delta _{1}\delta _{4}-\delta _{3}\delta _{2}\right) \left(
\beta _{1}\beta _{4}-\beta _{3}\beta _{2}\right) ,$

$a_{11}a_{13}-$ $a_{9}a_{15}=\allowbreak -\frac{1}{4}\alpha _{3}^{2}\delta
_{3}^{2}\left( \beta _{1}\beta _{4}-\beta _{3}\beta _{2}\right) \left(
-\gamma _{4}\gamma _{1}+\gamma _{2}\gamma _{3}\right) ,$

$a_{10}a_{12}-$ $a_{8}a_{14}=\allowbreak \frac{1}{4}\alpha _{3}^{2}\delta
_{1}^{2}\left( \beta _{1}\beta _{4}-\beta _{2}\beta _{3}\right) \left(
\gamma _{4}\gamma _{1}-\gamma _{2}\gamma _{3}\right) .$

From the above equations, we conclude the equalities of this criterion.

Next we argue that the inequalities hold. Given $\alpha $, $\beta $, $\gamma 
$ and $\delta $ are invertible. Assume that $a_{0}a_{3}=a_{1}a_{2},$ $%
a_{5}a_{6}=$\ $a_{4}a_{7}$, $a_{8}a_{11}=$\ $a_{9}a_{10}$ and $a_{13}a_{14}=$
$a_{12}a_{15}$.\ Then we obtain $\alpha _{1}\beta _{1}=0$, $\alpha _{1}\beta
_{3}=0$, $\alpha _{3}\beta _{1}=0$ and $\alpha _{3}\beta _{3}=0$.
Intuitively $\alpha _{1}\neq 0$ or $\alpha _{3}\neq 0$ results in $\beta
_{1}=\beta _{3}=0$. Therefore these inequalities hold.

\end{document}